\documentstyle[psfig]{l-aa}

\begin{document}

\def\ISO{{\it ISO}}
\def\COBE{{\it COBE}}
\def\IRAS{{\it IRAS}}
\def\etal{{\it et al.}}
\def\arcsec{\hbox{$^{\prime\prime}$}}
\def\simgt{\ga}
\def\simlt{\la}

\thesaurus{11     % galaxies
           (
           11.09.1 SMC; %Galaxies: individual:
           11.01.1;          % Galaxies: abundances
           09.04.1;          % ISM: dust,extinction
           13.09.1;          % Infrared: galaxies
           13.09.4       % Infrared: interstellar: lines
           )}
  
\title{Detection of mid-infrared Aromatic Hydrocarbon Emission Features from the Small Magellanic Cloud
\thanks{Based on observations with the Infrared Space Observatory (ISO).
 ISO is an ESA project with instruments funded by ESA Member States
   (especially the PI countries: France, Germany, the Netherlands and the
   United Kingdom) and with the participation of ISAS and NASA.}
}
%\subtitle{} 

\author{William T. Reach \inst{1,2} \and 
          Fran\c{c}ois Boulanger \inst{2} \and 
	  Alessandra Contursi \inst{1} \and
	  James Lequeux\inst{3}}
\offprints{W. T. Reach, reach@ipac.caltech.edu} 
 
\institute{Infrared Processing and Analysis Center, California Institute of Technology,
MS 100-22, Pasadena, CA 91125, USA
   \and
Institut d'Astrophysique Spatiale, B\^at.  121, 
 Universit\'e Paris XI, F--91405 Orsay cedex, France
   \and
Observatoire de Paris, 61 Av. de l'Observatoire, F-75014 Paris, France
%Service d'Astrophysique, CEA, DSM, DAPNIA, Centre d'Etudes de Saclay, 
%F-91191 Gif-sur-Yvette cedex, France   
}

\date{Received xxxx; Accepted xxxx} 

\maketitle
\markboth{W. T. Reach et al.: Hydrocarbons in the SMC}{ }

\begin{abstract}

The mid-infrared (5--16 $\mu$m) spectral energy distribution  for an individual
quiescent molecular cloud in the Small Magellanic Cloud (SMC) was observed using 
ISOCAM.
The spectrum is dominated by broad emission bands at 6.2, 7.7, and
11.3 $\mu$m, with  weaker bands at 8.6 and 12.7 $\mu$m. As these are the same
bands, with similar shape and relative strengths, as observed in the ISM 
of our
Galaxy, the same carriers must exist in both galaxies. The carriers are widely
consider to be  large molecules or clusters of aromatic
hydrocarbons, which absorb ultraviolet and visible photons and emit
mid-infrared photons during high-temperature pulses.  Based on the
brightness of the mid-infrared emission and the estimated strength of the
radiation field in the SMC, the absorption by aromatic hydrocarbons
 is of order 10\% of total dust
absorption, comparable to the case for Galactic dust. Ultraviolet observations
of extinction of most SMC stars have shown that dust in the SMC does not absorb in
the 2175 \AA\ feature that is so prominent in Milky Way extinction.  
If aromatic hydrocarbons and featureless extinction curves were ubiquitous in the SMC, 
then we would conclude that aromatic hydrocarbons are not the carriers of the 2175 \AA\ feature. 
However, SMC extinction curve measurements are biased toward  hot, luminous stars,
where aromatic hydrocarbons are destroyed, so that the absence of the 2175 \AA\ bump
may not be typical of SMC dust.
The presence of aromatic hydrocarbons
in the SMC further demonstrate that these molecules exist even in an
interstellar medium with an order-of-magnitude lower metallicity 
than in the disk of the Milky Way.

\def\extra{

The mid-infrared (5-16 micron) spectral energy distribution  for an individual
quiescent molecular cloud in the Small Magellanic Cloud (SMC) was observed
using  ISOCAM. The spectrum is dominated by broad emission bands at 6.2, 7.7,
and 11.3 microns, with  weaker bands at 8.6 and 12.7 microns. As these are the
same bands, with similar shape and relative strengths, as observed in the ISM 
of our Galaxy, the same carriers must exist in both galaxies. The carriers are
widely consider to be large molecules or clusters of aromatic hydrocarbons,
which absorb ultraviolet and visible photons and emit mid-infrared photons
during high-temperature pulses. Based on the brightness of the mid-infrared
emission and the estimated strength of the radiation field in the SMC, the
absorption by aromatic hydrocarbons is of order 10% of total dust absorption,
comparable to the case for Galactic dust. Ultraviolet observations of
extinction of most SMC stars have shown that dust in the SMC does not absorb in
the 2175 Angstrom feature that is so prominent in Milky Way extinction. If
aromatic hydrocarbons and featureless extinction curves were ubiquitous in the
SMC, then we would conclude that aromatic hydrocarbons are not the carriers of
the 2175  Angstrom feature. However, SMC extinction curve measurements are
biased toward hot, luminous stars, where aromatic hydrocarbons are destroyed,
so that the absence of the 2175 Angstrom bump may not be typical of SMC dust.
The presence of aromatic hydrocarbons in the SMC further demonstrate that these
molecules exist even in an interstellar medium with an order-of-magnitude lower
metallicity than in the disk of the Milky Way.

}


\end{abstract}

\section{Introduction}

The Small Magellanic Cloud (SMC) provides a unique environment
for studying interstellar matter. The abundance of C, N, O, and heavier elements
has been found to be an order of magnitude lower in the SMC
 than in the disk of the
Milky Way (\cite{dufour}), so heating, cooling, and chemical processes
dependent upon metal abundances can be expected to be different from our
local interstellar medium (ISM). 
%In addition, the density of star formation
%is currently higher in the SMC than in the Milky Way, so that the density of starlight
%is about an order of magnitude higher (\cite{Okumura}, \cite{rubioII}).
In particular, the properties of dust
in the SMC could be quite different from dust in the Milky Way, because
dust is composed almost entirely (by mass) of elements that are substantially
less abundant in the SMC. 
The visible extinction per unit H column density is about
an order of magnitude lower in the SMC than in the Milky Way,
suggesting that the total dust abundance may scale with the total metal 
abundance (\cite{Bouchet}).
The {\it spectrum} of SMC extinction (or the SMC `extinction curve') is significantly
different from that of our Galaxy: the SMC has relatively stronger ultraviolet
extinction, and it lacks the 2175 \AA\ bump that is so prominent in
the Milky Way extinction curve (\cite{savmath}). 
%The extinction curve depends
%on the dust properties as well as the environment (especially radiation field),
%so emission observations away from bright stars...

Infrared emission from the SMC was detected by the {\it Infrared Astronomical Satellite}
({\it IRAS}), with bright and extended 60 and 100 $\mu$m emission and weaker emission
at 12 and 25 $\mu$m (\cite{schwering}, \cite{sauvage}, \cite{Okumura}). The 12 $\mu$m {\it IRAS} image shows
very little diffuse emission, being dominated by a few compact regions. The weak
12 $\mu$m emission from the SMC suggested low-metallicity galaxies may be lacking 
the aromatic hydrocarbons that dominate the 12$\mu$m emission from the Milky Way.
In this paper, we describe one observation that is part of a larger
{\it Infrared Space Observatory} (\ISO; \cite{Kessler}) Guaranteed Time program 
designed to characterize the mid-infrared emission from interstellar dust
in a wide variety of environments.
The cloud SMC B1\#1 was selected as an extragalactic target to characterize
a quiescent, low-metallicity environment. The cloud was discovered serendipitously
during observations as part of the Swedish-ESO Submillimeter Telescope Key Program
to map CO-line emission from some star forming regions in the SMC 
(\cite{rubioII}; \cite{Lequeux}).
The cloud is far from any trace of O stars in existing H$\alpha$ and radio continuum maps
(see Fig.~5 of \cite{rubioII}),
and its molecular-line emission is narrow, indicating that the gas is dynamically
quiescent. Therefore the dust emission from the cloud is likely excited by a
combination of the average SMC radiation field and, possibly, some associated
stars. As such, the cloud represents a relatively simple laboratory
for studying the properties of SMC dust using its infrared spectrum, away from sites
of high-mass star formation.

\section{Observations}

Our observations were made with the mid-infrared camera, ISOCAM (\cite{CCesarsky}). 
The telescope was pointed toward J2000 coordinates 00$^h$45$^m$32.5$^s$, 
-73$^\circ$18$^\prime$46.3$^{\prime\prime}$ on 5 July 1996.
We used the 12\arcsec pixel-field-of-view lens, for which only the
central $3^\prime\times 3^\prime$ of the detector is illuminated.
A first spectrum (July 5, 1996) was obtained by rotating the circular-variable filters (CVF)
through the range 16.61 to 5.079 $\mu$m. 
%At each of 150 steps of the two CVFs, 12 frames of
%2.1 sec duration were taken; the total observing time was 1.25 hr.
A second spectrum (2 Oct 1997) was obtained by covering the same wavelengths but
rotating the CVFs both forward and backward.
During the second observation, we also observed a nearby reference position outside the 
SMC, at J2000 coordinates  00$^h$45$^m$32.5$^s$, -73$^\circ$18$^\prime$46.3$^{\prime\prime}$,
to measure the Milky Way emission in the 10.74--11.79 $\mu$m range.
For all observations, dark current was subtracted from the images using a 
library dark current image scaled such that it matches the level of the 
unilluminated edges of our images.
The brightness level after dark current subtraction was compared to that
expected from an interpolation of the {\it COBE} Diffuse Infrared Background
Experiment (DIRBE) broad-band observations at 4.9 and 12 $\mu$m. 
At low levels of illumination, the ISOCAM detectors exhibit some transient response,
and they take time to stabilize to the true sky brightness level. 
We applied a transient correction algorithm that takes into account the initial
rapid rise of the ISOCAM gain and a single exponential rise thereafter.
The transient corrections are only about 10\% for
wavelengths longer than 6 $\mu$m, because the change in brightness as the
CVF rotates is small; but
at the shortest wavelengths, the correction increases
to 60\%, because the sky brightness was decreasing
more quickly than the ISOCAM detector could stabilize. 
The images at each wavelength were corrected for zodiacal light,
stray light and vignetting by subtracting a special calibration observation 
of a blank field, scaled by a model spectrum of the zodiacal
light that matches the
\COBE/DIRBE data for the same position and date (Reach et al. 1996a,b).
To complement the CVF observations, we also present here a portion of an
image made with the ISOCAM through the LW2 filter (5--8.5 $\mu$m) on March 13, 1996.
This image was reduced using the CAM Interactive Analysis package and aligned
to stars in the Palomar Digital Sky Survey.

%For the second observation, only 6 frames per step were taken, but
%we rotated the CVFs from 16.61 to 5.079 $\mu$m and then back to 16.61 $\mu$m (to test
%the effect of detector gain variations). 
%At the reference position, we rotated the CVF from 11.79 to 10.74 $\mu$m and
%back again.

%The accumulated charge density in the detectors is proportional to the observed
%brightness minus the dark current and divided by the (electronically-applied) gain;
%this quantity varies from 1 to 18 ADU during the observation described here.
% (\cite{abergel}).
%Because both the transient
%and dark current corrections are rather large for
%the short wavelengths, the brightness observed at wavelengths shorter than 6 $\mu$m
%is not reliable. At longer wavelengths, however, neither the transient correction nor the
%dark current offset were significant, and the absolute brightness levels are
%reliable.

\section{Results}

\begin{figure}
%\picplace{9cm}
\psfig{figure=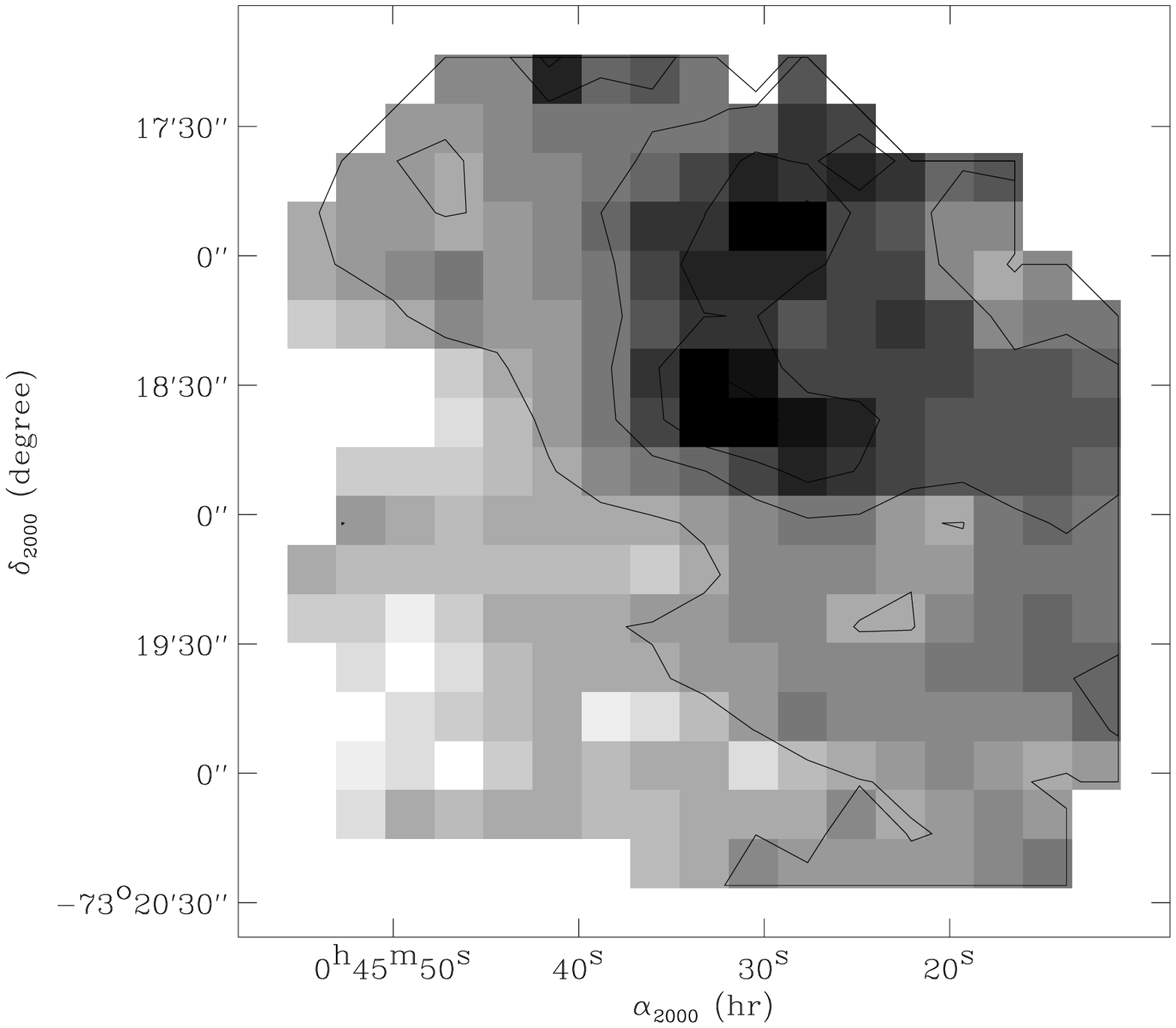,height=9cm,angle=90}
\psfig{figure=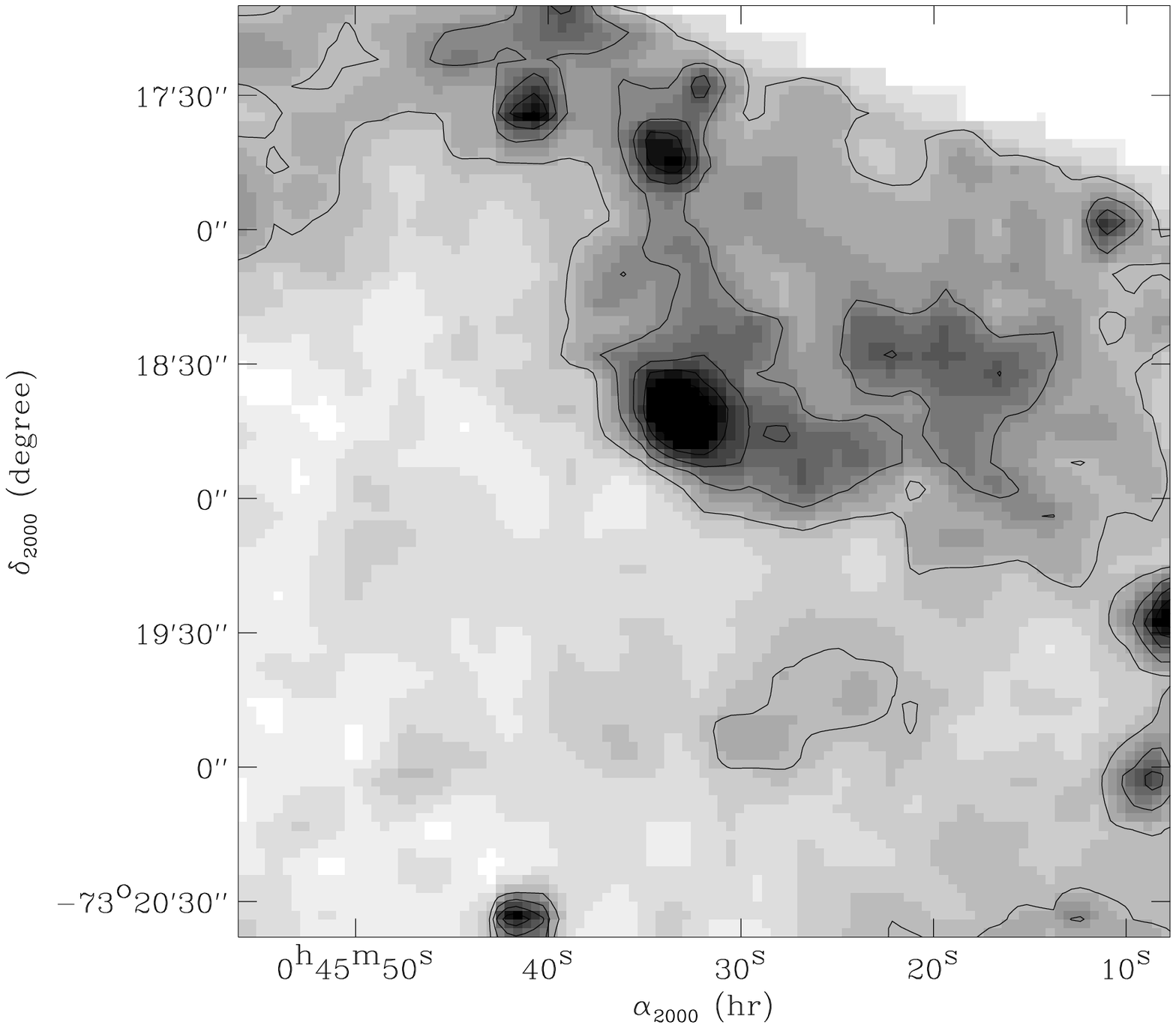,height=9cm,angle=90}
\caption{Images of the cloud SMC B1\#1 from ISOCAM. {\it (top)} Image
through the circular-variable-filter, within the 11.3 $\mu$m 
aromatic hydrocarbon feature, with $12^{\prime\prime}$ pixels.
The image shows the average brightness from 11.16--11.48 $\mu$m wavelength,
minus the average brightness from in two windows just outside the feature
(10.52--10.84 and 11.79--12.1 $\mu$m). 
Contours are drawn at 0.5, 1, 1.5, and 2 MJy~sr$^{-1}$.
This image is dominated by diffuse emission from the SMCB1\#1 molecular cloud.
{\it (bottom}) Image through the wide, LW2 filter (5--8.5 $\mu$m),
with $6^{\prime\prime}$ pixels. 
Contours are drawn at 0.3, 0.6, 0.9, and 1.2 MJy~sr$^{-1}$.
The image contains diffuse emission from
the SMCB1\#1 molecular cloud as well as unrelated SMC stars and the
embedded source SMC004532.2-731840.  
}
\label{smcmap}
\end{figure}

\begin{figure}
%\picplace{9cm}
\psfig{figure=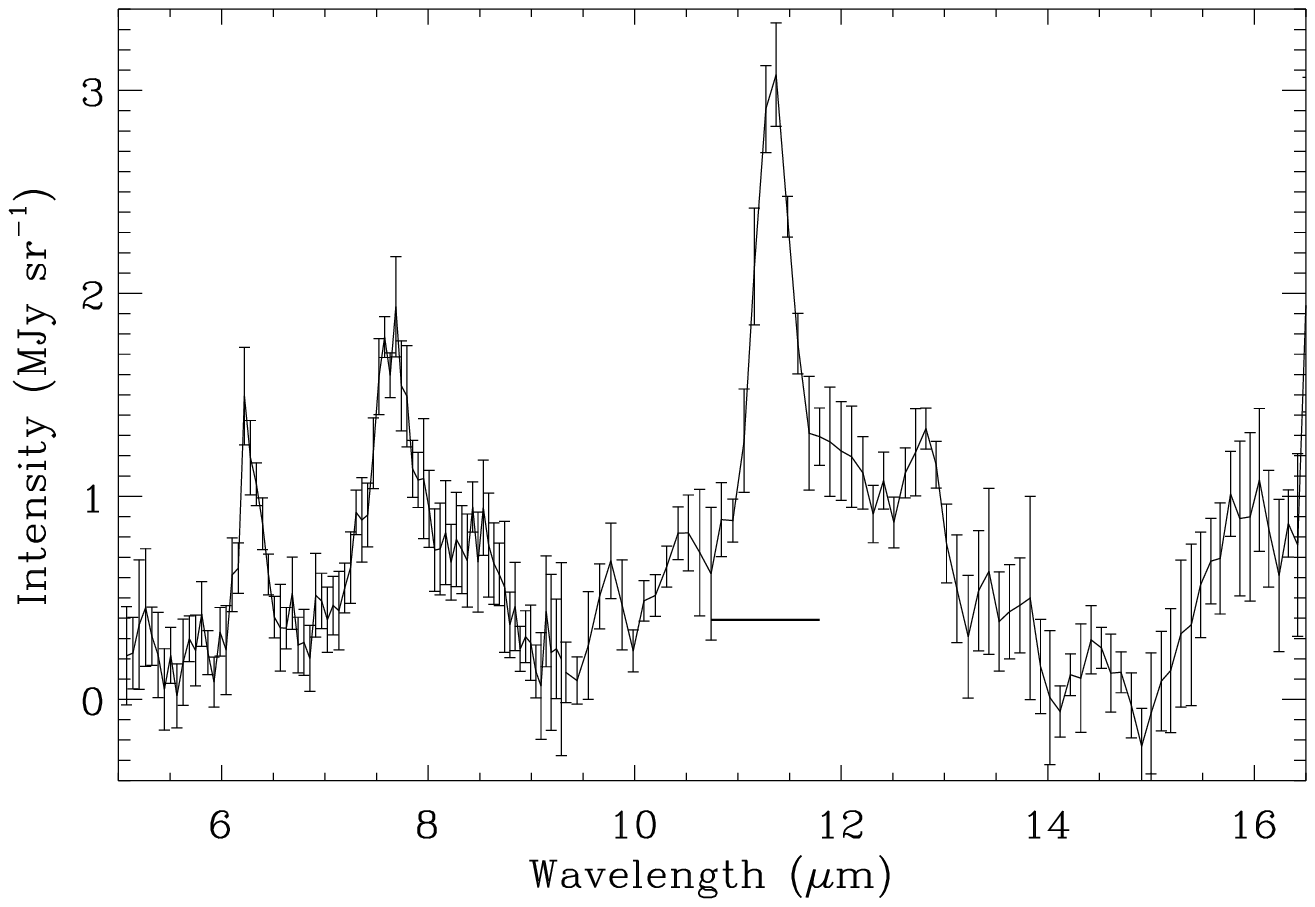,width=9cm}
\psfig{figure=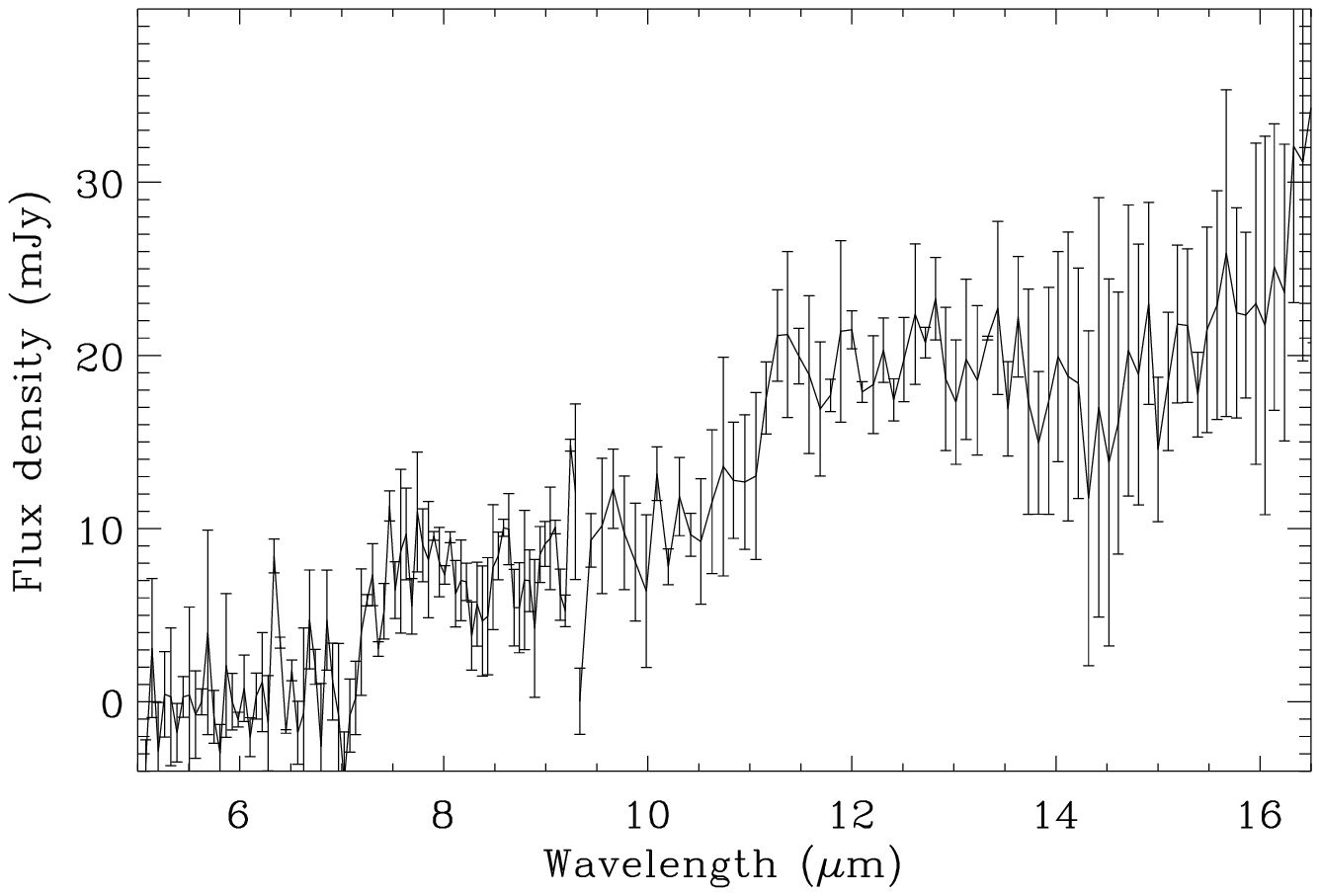,width=9cm}
\caption{
{\it (a)} Mid-infrared spectrum of the surface brightness of the cloud SMC B1\#1. 
This spectrum includes only a
$7\times 7$ pixel region with emission from the
molecular cloud, excluding the $3\times 3$ pixel region around the point-like
embedded infrared source. The error bars represent the random fluctuations plus
the systematic differences between the spectra taken at three different times.
The horizontal bar under the 11.3 $\mu$m feature is an upper limit to the Milky
Way contribution to the feature, obtained 
using the same CVF wavelengths but offset spatially from the SMC.
{\it (b)} Spectrum of the mid-infrared flux density embedded source 
SMC004532.2-731840, made from the brightest pixel in the CVF continuum image.}
\label{smcspec}
\end{figure}

%The ISOCAM images reveal an embedded source within a region of diffuse emission.
Figure~\ref{smcmap}a shows an image of the region we observed, in 
the 11.3 $\mu$m aromatic hydrocarbon feature.
The  11.3 $\mu$m emission feature comes mostly from the
northwest corner of the image.
Figure~\ref{smcmap}b shows an image in the broad LW2 filter (5--8.5 $\mu$m).
The LW2 image shows some diffuse emission that is spatially associated with the 
emission in the 11.3 $\mu$m feature, as well as several point sources. 
The similarity between the wide-band image and the 11.3 $\mu$m image suggests that
they are both tracing the same type of emission.

The diffuse emission in the 11.3 $\mu$m image 
is coincident with the molecular cloud SMC B1\#1. 
This cloud has been mapped in 
the CO(2--1) emission line, revealing a cloud size of
50\arcsec$\times$40\arcsec (\cite{rubioV}), similar in size and shape
to the 11.3 $\mu$m emission we see with ISOCAM.
Figure ~\ref{smcspec}(a) shows the spectrum of the cloud, averaged over
the diffuse emission, comparable to the extent of the CO emission, and
excluding the embedded source, which is the peak pixel in the CVF image
and a point-like source in the LW2 filter image.
It is evident that the diffuse emission
is dominated by features at 6.2, 7.7, and 11.3 $\mu$m,
corresponding to the brightest of the mid-infrared features
that appear in similar spectra of Galactic interstellar clouds
({\it e.g.} \cite{Rophpap}). 
The reasonably good correlation between the diffuse emission in
the LW2 image and the 11.3 $\mu$m feature image is most likely due to the LW2 filter
containing the 6.2 and 7.7 $\mu$m features. Thus 
Fig.~\ref{smcmap}(a) shows only emission from one aromatic hydrocarbon
 feature, while
Fig.~\ref{smcmap}(b) shows a combination of point-like
continuum sources and diffuse emission from aromatic hydrocarbon features.

The Milky Way foreground is faint compared to our target SMC B1\#1:
no aromatic hydrocarbon features (from the Milky Way {\it or} the SMC), are detected in the 
southeast corner of our CVF image, nor in the reference position 
outside the SMC (Fig.~\ref{smcspec}).
Thus it is evident that the carrier of these features, widely 
considered to be aromatic hydrocarbons, also exist in the SMC.

The relatively bright point source in the LW2 image,
SMC004532.2-731840, has an
extremely red spectrum, shown in Fig.~\ref{smcspec}(b),
rising continuously from 5--16 $\mu$m.
The source spectrum is relatively featureless, so it does not appear as a 
prominent source in the 11.3 $\mu$m feature image. 
Using the filter images (with $6^{\prime\prime}$ pixels), the point-like
source has a flux densities of 23 and 5 mJy in the LW3 (12--18 $\mu$m) and
LW2 (5--8.5 $\mu$m) filters, respectively. 
The source's luminosity in the mid-infrared, integrating over 5--16 $\mu$m, is
$\sim 300$ L$_\odot$.
A possible Galactic analog for this source is an ultracompact H~II region:
the mid-infrared spectrum is similar to that observed for an ultracompact H~II
region in M~17 (\cite{DCesarsky}). The source could also be a reflection
nebula, but the featureless and extremely red
spectrum is completely different from reflection nebulae around Milky Way stars.

%The expected Galactic foreground, using the Galactic H~I column density
%(\cite{cleary,schwering}), infrared colors (\cite{arendt,dwek97}), and mid-infrared spectrum
%(\cite{Rophpap})
%scaling laws observed for the galactic interstellar medium:
%$I_{12}/I_{100}=0.046$ (\cite{dwek97}), $I_{100}/N_{\rm H}=0.6\times 10^{-20}$ 
%MJy sr$^{-1}$ cm$^{2}$ (\cite{arendt}), and the Galactic H~I column density
%toward the SMC which is $4\times 10^{20}$ averaged over a half-degree beam (\cite{cleary}).
%The Galactic foreground toward the SMC is is 180 times fainter than the emission from
%is only 0.27 MJy sr$^{-1}$ at the peak of the 11.3 $\mu$m emission feature, more
%than an order of magnitude weaker than we observe for the SMC diffuse cloud.

\section{Discussion}

\noindent{\it Relation to extinction curve---}
The SMC is one of the most unusual environments in which to observe 
aromatic hydrocarbons in abundance
because of two well-established facts. First, the metallicity of the SMC is lower
than in our Galaxy by a factor of 10, and the C/O ratio is also lower than in our Galaxy,
which could affect the abundance of aromatic hydrocarbon
 molecules, which composed almost entirely of C by mass (\cite{dufour}).
Second, the measured extinction curve of SMC dust is significantly different from
that of Galactic dust. In particular, most SMC extinction curves lack the prominent 
2175 \AA\ bump 
so obvious in the Galactic extinction curve (\cite{savmath}). This bump is widely attributed
to graphitic particles. Clusters of aromatic hydrocarbons 
may have similar spectral properties to graphite. 
Because the 2175 \AA\
feature is missing from the SMC, while the aromatic hydrocarbon
 abundance is high in the SMC, our results suggest
that aromatic hydrocarbons are not responsible for the 2175 \AA\ bump in the Milky Way.
However, at least one SMC star, Sk 143, has an extinction curve 
with the same shape as that of the Milky Way, which suggests strong regional
variations in the SMC extinction curve (\cite{lequeux}, \cite{gordon}). 
Also, we have only detected the aromatic features from one quiescent cloud, which
may or may not be representative of typical SMC material.
Therefore, it is possible that we could be comparing the 
UV-visible extinction and mid-infrared emission for regions
with different dust properties.
This problem could be eliminated in the future,
by observing the extinction curve through
the same type of region from which mid-infrared emission in observed.
The aromatic hydrocarbons must absorb a substantial fraction (see below) of the interstellar
radiation field, but we cannot be sure what photon energy range is exciting them.
Observations of reflection nebulae excited by stars with a range of spectral
types indicate that aromatic hydrocarbons may not require
far-ultraviolet photons for excitation (\cite{uchida}).
%Therefore, as a corollary derived from the observed high aromatic hydrocarbon
% abundance in the SMC, our results suggest
%that aromatic hydrocarbons
% are not solely responsible for the fast far-ultraviolet 
%rise of the extinction in the SMC.

\begin{table}
\caption[]{Lorentzian fit to SMC Infrared Emission Features}\label{fittab}
\begin{flushleft}
\begin{tabular}{cccc}
\hline
 \multicolumn{1}{c}{$I_\nu(0)$} & 
\multicolumn{1}{c}{$\lambda_0$} & \multicolumn{1}{c}{FWHM} & \multicolumn{1}{c}{$I$} \\ 
\multicolumn{1}{c}{(MJy sr$^{-1}$)} & 
\multicolumn{1}{c}{($\mu$m)} & \multicolumn{1}{c}{($\mu$m)} & 
\multicolumn{1}{c}{(nW m$^{-2}$ sr$^{-1}$)} \\
\hline
$ 1.17\pm 0.09$ & $ 6.26\pm 0.01$ & $ 0.20\pm 0.01$ & $14.4\pm 1.3$ \\
$ 1.56\pm 0.05$ & $ 7.65\pm 0.01$ & $ 0.53\pm 0.01$ & $33.0\pm 1.4$ \\
$ 0.48\pm 0.06$ & $ 8.48\pm 0.02$ & $ 0.36\pm 0.04$ & $ 5.5\pm 0.9$ \\
$ 2.71\pm 0.07$ & $11.34\pm 0.01$ & $ 0.54\pm 0.01$ & $26.7\pm 0.9$ \\
$ 0.85\pm 0.05$ & $12.59\pm 0.03$ & $ 0.90\pm 0.05$ & $11.4\pm 1.0$ \\
\hline
\end{tabular}
\end{flushleft} 
\end{table} 

\noindent{\it Comparison to Milky Way spectra---}
The mid-infrared spectrum of the SMC contains features at the same
wavelengths as most galactic sources; however, the feature-to-feature
ratios are quite distinct for the SMC. The spectrum of the diffuse
cloud emission is well fit by a
sum of 4 Lorentzians, representing the 6.2, 7.7, 11.3, and 12.6 $\mu$m
features. A Lorentzian shape is a better fit and more physically
justified than a Gaussian fit (Boulanger et al. 1998). 
Other than the wings of the Lorentzians and a broad pedestal
under the 11.3 and 12.6 $\mu$m features, no continuum
was detected toward the diffuse cloud (within the uncertainties
The relative strengths of the emission features are significantly
different from those of Milky Way objects.
Compared to the observations of H~II
regions, reflection nebulae, and diffuse clouds,
summarized by Lu (1998),
the ratio of (11.3)/(7.7) features is higher in SMCB1\#1.
Compared to the B-star excited emission from the $\rho$ Oph molecular clouds
(Boulanger et al. 1996, 1998),
the ratio of (11.3)/(7.7) features is 3 times higher in SMCB1\#1.
%The observed combination of high (6.2)/(7.7) and high (11.3)/(7.7) in
%the SMC is counter to the tentative trend in these ratios seen in
%Milky Way spectra (\cite{lu98}).
%We suspect that the structure and size distribution of the carriers
%of the emission features are somewhat different in the SMC than they are in
%the Milky Way. 
Ionization of polycyclic aromatic hydrocarbon molecules
(PAH) enhances the 6--9 $\mu$m features
relative to the 11--14 $\mu$m features (\cite{joblin96,alla99}), a
signature that is opposite to our observations for the SMC. 
If the aromatic hydrocarbons in the $\rho$ Oph region are ionized, then
the comparison would suggest that the aromatic hydrocarbons
 in SMC B1\#1 are {\it more
neutral}. We suggest that another explanation may be more likely.
The (11.3)/(7.7) feature ratio is proportional to the fraction of 
emission arising from C--H bonds as opposed to C--C bonds.
The high value of this ratio in the SMC suggests there are relatively
more C--H bonds. This seems consistent with the aromatic
hydrocarbons forming in a more
reducing environment (higher H/C abundance ratio) in the SMC, 
where the abundance of C in the gas from which 
the molecules form is a factor of 10 lower than in the Milky Way.
If the hydrocarbons in the SMC are indeed more hydrogenated than those in
the Milky Way, we predict a relatively bright 3.3 $\mu$m line from
diffuse SMC gas. 
%(On the other hand, if the aromatic 
%hydrocarbons in the SMC are less 
%ionized than in the Milky Way, then the 3.3 $\mu$m feature may be 
%relatively weak, if
%we use Fig.~2 of Allamandola et al. as a guide.)
Our current understanding the nature of interstellar
aromatic hydrocarbons is still in its
early stages because of the complex inter-relationship between
the excitation, chemistry, collisional destruction, and
ionization. By providing a new aromatic hydrocarbon
 spectrum from a unique environment,
the observations reported here should help toward disentangling
the various processes that shape the aromatic hydrocarbon
 features and their carriers.

\begin{figure}
\psfig{figure=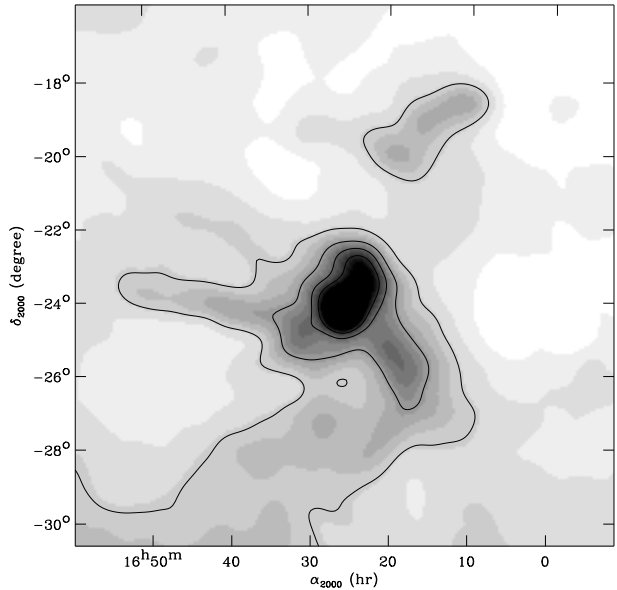,height=9cm,angle=90}
\caption{IRAS Image of the $\rho$ Oph region, as it would appear if it
were observed at the distance of the SMC with $6^{\prime\prime}$ resolution.
The bright and unrelated star Antares was deleted from the original IRAS 
map before projecting the map to the SMC distance.
Contours are drawn at 2 MJy sr$^{-1}$ intervals of IRAS 12 $\mu$m
surface brightness.
The brightest pixel in the map has an IRAS 12 $\mu$m band surface brightness
of 22 MJy sr$^{-1}$, while the periphery of the main cloud and the adjacent
clouds in the image have brightnesses of a few MJy~sr$^{-1}$, comparable
to the SMC B1\#1 cloud that we observed with ISOCAM.}
\label{rophmap}
\end{figure}

\noindent{\it Comparison to Milky Way nebulae---}
We can put SMC B1\#1 into context by comparing
it to what known galactic nebulae would look like at the distance
of the SMC. To this end, we used the {\it IRAS ISSA} (\cite{issaref})
to make simulated maps of three galactic complexes:
(1) the O-star-excited H~II region around the Orion Trapezium,
(2) the B-star-excited reflection nebula around $\rho$ Oph, and
(3) the star-forming complex in Taurus.
The physical sizes of all three nebulae are in the same order of magnitude
as SMC B1\#1 ($\sim 10$ pc). The 12 $\mu$m surface brightnesses at
1.7 pc physical resolution ($12^{\prime\prime}$ at SMC distance) of the 
objects are 350, 12, and $< 0.5$  MJy sr$^{-1}$, respectively;
for comparison, the brightness of SMC B1\#1 is about 2 MJy~sr$^{-1}$.
It is clear that SMC B1\#1 is significantly
different from the Trapezium, both because it is much less bright in
the mid-infrared, and because it has no associated radio or optical 
H~II region. It is also evident that SMC B1\#1 is different 
from the Taurus clouds, because it is brighter in the mid-infrared.
The best Galactic analog for SMC B1\#1 is $\rho$ Oph.
The smoothed image of $\rho$ Oph, shown in Fig.~\ref{rophmap},
is comparable to that of SMC B1\#1, with 
one bright point source and diffuse emission on a 10~pc size scale.
The point-source in our ISOCAM image of SMC B1\#1 could be 
a reflection nebula, analogous to the bright mid-infrared
region in $\rho$ Oph imaged by ISOCAM (\cite{abergel}), but with a 
rather different spectrum; or it could be an ultracompact H~II region
from a highly embedded late-type O star.
% Follow-up 
%observations are needed to determine which stars contribute to the dust
%heating in this region. 
%A good analog for SMC B1\#1 is probably the
%cloud just to the northwest of $\rho Oph$, in the upper left of 
%Fig.~\ref{rophmap}, because it has the same surface brightness when
%smoothed to the same resolution.

\noindent{\it Estimated abundance of aromatic hydrocarbons---}
The abundance of aromatic hydrocarbons
 in the SMC B1\#1 cloud can be roughly estimated from the
brightness of the infrared features in the diffuse emission from
SMC B1\#1, where the dust heating is due to the
diffuse SMC interstellar radiation field. 
The total energy absorbed by a completely dark cloud is the integrated energy from the
incident interstellar radiation field.
The fraction of the radiation field that is absorbed by aromatic hydrocarbons
 can be calculated from the observed infrared brightness as
\[ f_{AH} = \frac{I_{AH}}{\chi I_{ISRF} g} \]
where $I_{AH}$ is the intensity of the observed aromatic hydrocarbon features,
$g$ is the fraction of all aromatic hydrocarbon emission that comes out in the
observed features, and $I_{ISRF}$ is the intensity of the ultraviolet
Solar Neighborhood radiation field (\cite{mmp}).
In practice, the value of $f_{AH}$ obtained from this equation is a lower limit,
because the cloud surface may not be completely dark and may not fill the beam
uniformly.
The radiation field of the observed cloud is $\chi$,
in units of $I_{ISRF}$.
Adding together the 1620~\AA\ fluxes of the stars in the H~II regions seen with 
the {\it UIT} (\cite{cornett}), and including the estimated fluxes of the H~II
regions DEM 21, 28 and 16 based on their H$\alpha$ brightness, we estimate that 
the radiation field at the location of SMC B1\#1 has $\chi\sim 3$, with an uncertainty
of about a factor of 2.
%In the SMC~B region, which is relatively
%quiet, we assume $\chi\sim 1$--10.
If the exciting star of the point-like reflection nebula in SMC B1\#1
is a B3 or later star, its flux would exceed that of the radiation
field only within $< 3$~pc of the star (or less, if extinction were
included), consistent with the unresolved
source in our image but of little or no importance for the diffuse emission.
The ISOCAM wavelength range includes essentially all of the aromatic hydrocarbon
emission, so $g\simeq 1$,
although it is possible that we miss some emission in a quasi-continuum
under the features or outside of our observed wavelength range.
Using the sum of the feature brightness from Table~\ref{fittab},
we find $f_{AH}> 0.05$.

%A similar estimate for the Milky Way, using $\chi=1$ and the typical 
%12 $\mu$m brightness of a dark cloud, yields ??????
%COMPARE TO MILKY WAY

To estimate directly the energy emitted by large dust grains 
in the SMC, we use archival ISOPHOT (\cite{phtref}) observations. An image 
of the SMC-B1 region, covering $12^\prime\times 12^\prime$ including the
region of our ISOCAM observation, was made at 135 $\mu$m wavelength 
with the C2 detectors.
The far-infrared emission is faint at the location of our ISOCAM image,
with no evidence of a far-infrared peak corresponding to the 
mid-infrared peak.
There is some far-infrared emission extending north of our target region, 
suggesting that SMC B1\#1 is a clump at the southern edge of a 
large, cold cloud. The far-infrared surface brightness at the location of
our mid-infrared peak SMC~B1\#1 is $\simlt 10$ MJy~sr$^{-1}$, measured with
respect to nearby dark pixels in the ISOPHOT image.
If the far-infrared spectrum is typical of large dust grains in the Milky Way
(\cite{boul}, \cite{dwek97}), then
the integrated far-infrared emission is 2 times brighter than $\nu I_\nu$
at 135 $\mu$m, so the far-infrared surface brightness of our cloud is 
$<460$ nW m$^{-2}$ sr$^{-1}$.
Comparing to the sum of the aromatic hydrocarbon feature brightnesses in Table~\ref{fittab},
the relative amount of energy absorbed by aromatic hydrocarbon
 as compared to big grains
is $I_{AH}/I_{BG} \simgt 0.2$.
This is comparable to the fraction of energy absorbed
by aromatic hydrocarbons
 in our Galaxy (\cite{dwek97}). It is also consistent with our abundance
estimate in the previous paragraph.

\section{Conclusions}

Using ISOCAM observations of a molecular cloud in the SMC, we found
that the mid-infrared emission of its low-metallicity interstellar medium
is dominated by broad aromatic hydrocarbon features at 6.2, 7.7, 8.6,
11.3, and 12.7 $\mu$m. The locations of the features are similar to those
found in the Milky Way interstellar medium, but the brightness ratios
among the features are different. The 11.3 $\mu$m
feature is relatively brighter in the SMC, suggesting 
relatively more emission from C--H bonds compared to C--C bonds,
which we can explain as a difference in the aromatic hydrocarbons
forming in a lower metallicity (higher H/C abundance ratio) environment in the SMC.
The abundance of aromatic hydrocarbons, 
relative to the dust that makes the rest of the 
absorption, is similar in the SMC B1\#1 cloud and typical Milky Way dark clouds.
While the properties, and the total abundance, of dust in the SMC 
and Milky Way differ significantly, the aromatic hydrocarbons
 and larger grains
are likely to play similar roles in the interstellar media of the two 
galaxies.

Previous studies of the mid-infrared emission from SMC suggested that its dust 
lacks the 2175 \AA\ feature and may
be deficient in aromatic hydrocarbons. 
These effects may both be due to the fact that massive star forming regions 
dominate the infrared emission and provide the beacons toward which extinction
curves are measured. 
Aromatic hydrocarbons are destroyed in such high-radiation environments (\cite{ryter}). 
Our results suggest that aromatic hydrocarbons exist in the SMC with an abundance (relative
to that of large grains) comparable to that in the Milky Way.
These results probably apply to other low-metallicity and high-redshift galaxies.

\acknowledgements
We thank Nanyao Lu for his help in processing and understanding the ISOPHOT
observations that were used in this paper.


\begin{thebibliography}{}

\bibitem[Abergel et al. 1996]{abergel}
Abergel, A. et al. 1996, A\& A, 315, L329

\bibitem[Allamandola, Hudgins, \& Sanford 1999]{alla99} Allamandola, L. J.,
Hudgins, D. M., \& Sanford, S. A. 1999, ApJL, 411, L115

\bibitem[Arendt et al. 1998]{arendt}
Arendt, R. G., et al. 1998, ApJ, 508, 74

\bibitem[Bouchet et al. 1985]{Bouchet}
Bouchet, P., Lequeux, J., Maurice, E., Pr\'evot, L.,  
Pr\'evot-Burnichon, M. L. 1985, A\&A, 149, 330

\bibitem[Boulanger, et al. 1988]{boul} 
Boulanger, F., Beichman, C., Desert, F. X., Helou, G., Perault, M. and Ryter, C. 1988, 
ApJ, 332, 328 

\bibitem[Boulanger et al. 1996]{Rophpap}
Boulanger, F. et al. 1996, A\&A, 315, L325 

\bibitem[Boulanger et al. 1998]{lorentzpap}
Boulanger, F., Boissel, P., Cesarsky, D., \& Ryter, C. 1998, A\&A, 339, 194

\bibitem[Cesarsky, C. et al. 1996]{CCesarsky}
Cesarsky, C. et al. 1996, A\&A, 315, 32 

\bibitem[Cesarsky, D. et al. 1996]{DCesarsky}
Cesarsky, D., Lequeux, J., Abergel, A., Perault, M., Palazzi, E.,
Madden, S., \& Tran, D. 1996, A\&A, 315, L309

\bibitem[Cleary, Haslam, \& Heiles 1979]{cleary}
Cleary, M. N., Haslam, C. G. T., \& Heiles, C. 1979, A\& AS, 36, 95

\bibitem[Contursi et al. 1998]{contursi}
Contursi, A., et al. 
1998, A\& A, 336, 662
% interprets ISOCAM images of LMC as being due to PAH

\bibitem[Cornett et al. 1997]{cornett}
Cornett et. al 1997, AJ 113, 1011 

\bibitem[Dwek et al. 1997]{dwek97}
Dwek, E., et al.
%Arendt, R. G., Fixsen, D. J., Sodroski, T. J., Odegard, N.,
%Weiland, J. L., Reach, W. T., Hauser, M. G., Kelsall, T., Moseley, S. H., 
%Silverberg, R. F., Shafer, R. A., Ballester, J., Bazell, D., \& Isaacman, R. 
1997, ApJ, 475, 565

\bibitem[Dufour et al. 1982]{dufour}
Dufour, R. J., Shields, G. A., Talbot, R. J. 1982, ApJ, 252, 461

\bibitem[      and Clayton 1998]{gordon}
Gordon, K. A., \& Clayton, G. C. 1998, ApJ, 500, 816

\bibitem[Israel et al. 1993]{Israel}
Israel, F. P. et al. 1993, A\&A, 276, 25

\bibitem[Joblin et al. 1996]{joblin96}
Joblin, C., Tielens, A. G. G. M., Geballe, T. R., \& Wooden, D. H.
1996, ApJL, 460, L119
% interprets line ratio 8.6/11.3 variations with distance from central star
% as due to variation in PAH ionization, because 6-9 um [CC and CH in plane]
% features much enhanced for cation whereas 11-14 [CH out of plane] are not

\bibitem[Kessler et al. 1996]{Kessler}
Kessler, M. et al. 1996, A\&A, 315, 27

\bibitem[Lemke et al. 1996]{phtref} Lemke, D., et al. 1996, A\& A, 315, L64

\bibitem[Lequeux et al. 1994]{Lequeux}
Lequeux, J., Le Bourlot, J., Pineau des For\^{e}ts, G., Roueff, E., 
Boulanger, F., Rubio, M. 1994, A\&A, 292, 371

\bibitem[Lequeux et al. 1982]{lequeux}
Lequeux, J., Maurice, E., Pr/'evot-Burnichon, M.-L., Pr\'evot, L., \&
Rocca-Volmerange, B. 1982, A\& A, 113, L15

\bibitem[Lu 1998]{lu98}
Lu, N. 1998, ApJL, 498, L65
% systematic tabulation of galactic PAH line ratios

\bibitem[Mathis, Mezger, \& Panagia 1983]{mmp}
Mathis, J. S., Mezger, P. M., Panagia, N. 1983, A\&A, 128, 212

\bibitem[Okumura 1993]{Okumura}
Okumura, K. 1993, Ph. D. Thesis, Universit\'e Paris Sud

\bibitem[Reach et al. 1996a]{DIRBEpap}
Reach, W. T. et al. 1996a, in {\it Unveiling the Cosmic Infrared Background}, ed. E. Dwek
(AIP: New York), p. 37

\bibitem[Reach et al. 1996b]{Zodypap}
Reach, W. T. et al. 1996b, A\&A, 315, L381

\bibitem[Rubio et al. 1993]{rubioII}
Rubio, M. et al. 1993, A\&A, 271, 1

\bibitem[Rubio et al. 1996]{rubioV}
Rubio, M. et al. 1996, A\&AS, 118, 263

\bibitem[Ryter, Puget and Perault 1987]{ryter} Ryter, C., 
Puget, J. L. and Perault, M. 1987, A\& A, 186, 312 

\bibitem[Sauvage et al. 1990]{sauvage} 
Sauvage, M., Thuan, T. X., \& Vigroux, L. 1990, A\&A, 237, 296

\bibitem[Savage \& Mathis 1979]{savmath}
Savage, B. D., \& Mathis, J. S. 1979, ARA\& A, 17, 73

\bibitem[Schwering \& Israel 1991]{schwering}
Schwering, P. B. W., \& Israel, F. P. 1991, A\& A, 246, 231

\bibitem[Uchida, Sellgren, \& Werner 1998]{uchida}
Uchida, K. I., Sellgren, K., \& Werner, M. 1998, ApJ, 493, L109

\bibitem[Wheelock et al. 1994]{issaref}
Wheelock S. L., Gautier, T. N., Chillemi, J., Kester, 
D., McCallon, H., Oken, C., White, J., Gregorich, D., Boulanger, F., 
and J. Good 1994, {\it IRAS Sky Survey Atlas: Explanatory Supplement}.
(JPL/Caltech: Pasadena)



\end{thebibliography}
\end{document}